\documentclass[11pt]{article}

\advance \textheight by \topskip
\makeatletter
\@addtoreset{equation}{section}
    
\makeatother

\usepackage{amsmath,amssymb}

\begin{document}

\title{
{\bf   On  hidden broken nonlinear
 superconformal  symmetry of
conformal mechanics and nature of double nonlinear
superconformal
symmetry }}

\author
{ {\sf Francisco Correa${}^a$}\thanks{E-mail:
francisco.correa@usach.cl}, {\sf Mariano A. del
Olmo${}^b$}\thanks{E-mail: olmo@fta.uva.es}
 {\sf\ and Mikhail S. Plyushchay${}^a$}\thanks{E-mail:
mplyushc@lauca.usach.cl}
\\[4pt]
{\small \it ${}^a$Departamento de F\'{\i}sica,
Universidad de Santiago de Chile}\\
{\small \it Casilla 307, Santiago 2, Chile}\\
{\small \it ${}^b$Departamento de F\'{\i}sica,
Universidad de Valladolid}\\
{\small \it E-47011, Valladolid, Spain}\\
}
\date{}

\maketitle

\begin{abstract}
We show that for positive integer  values $l$ of the
parameter in
the conformal mechanics model the system possesses a hidden
 nonlinear superconformal symmetry, in which
reflection plays a role of the grading operator. In addition
to
the even $so(1,2)\oplus u(1)$-generators, the superalgebra
includes $2l+1$ odd integrals, which form the pair of
spin-$(l+\frac{1}{2})$ representations of the bosonic
subalgebra
and anticommute for  order $2l+1$ polynomials of the even
generators. This hidden symmetry, however, is broken at the
level
of the states in such a way that the action of the odd
generators
violates the boundary condition at the origin. In the
earlier
observed double nonlinear superconformal symmetry, arising
in the
superconformal mechanics for certain values of the boson-
fermion
coupling constant, the higher order symmetry is of the same,
broken nature.
\end{abstract}

\section{Introduction}
The conformal mechanics model of De Alfaro, Fubini and
Furlan
\cite{AFF,IKL} is the simplest nontrivial (0+1)-dimensional
conformal field theory. The interest to it and to its
supersymmetric extension \cite{AP,FR,IKL2} has revived
recently in
the context of the black hole physics, AdS/CFT
correspondence and
integrable Calogero-Moser type systems
\cite{BH}--\cite{IKLe}.
This model provides an example of quantum mechanical system,
in
which the problem of a self-adjointness of the Hamiltonian
arises
in certain region of its parameter values
\cite{Me,Narn,Basu,Falom1}. The latter aspect is related to
the
problem of existence of bound quantum states and spontaneous
breaking of scale symmetry, that also finds applications  in
the
physics of black holes \cite{Govi}.

Recently, it was observed  that a certain change of the
boson-fermion coupling constant in the supersymmetric
conformal
mechanics model gives rise to a radical change of symmetry:
instead of the $osp(2|2)$ superconformal symmetry, the
modified
system is characterized by its nonlinear generalization
\cite{LeiP,AnaP}. It was also found that when the boson-
fermion
coupling constant takes integer values, the system is
described
simultaneously by the two nonlinear superconformal
symmetries of
the orders relatively shifted in odd number \cite{LeiP}. In
particular, such nonlinear superconformal symmetry arises in
the
original superconformal mechanics model \cite{AP,FR} in
addition
to the $osp(2|2)$. However, the nature of this  double
superconformal symmetry left to be mysterious.

In the present Letter we show that when the parameter $\alpha$ of
the purely bosonic conformal mechanics model (\ref{Hconf}) takes
integer values, in addition to the $so(1,2)$ conformal symmetry the
system possesses a set of the integrals of motion which are odd
differential operators. Identifying the nonlocal reflection operator
as a grading operator, we find that these additional integrals
extend the $so(1,2)$ to the nonlinear superconformal symmetry
discussed in \cite{LeiP,AnaP}. However, this hidden nonlinear
superconformal symmetry of the conformal mechanics model is broken
at the level of the states. Similarly to the usual spontaneous
supersymmetry breaking mechanism, where a zero energy state loses
the normalizability due to a violation of the boundary condition at
infinity, here the odd generators acting on the Hamiltonian
eigenstates (being the non-normalizable, scattering states) produce
its other eigenstates, which violate the boundary condition at the
origin. Then, we show that in the case of the double superconformal
symmetry observed in ref. \cite{LeiP}, the symmetry of higher order
has the same, broken nature.

The Letter is organized as follows. In Section 2 we show that a free
particle on a line possesses a hidden $osp(2|2)$ superconformal
symmetry in the exact, unbroken phase, in which reflection plays a
role of a grading operator. Section 3 is devoted to the discussion
of a hidden broken nonlinear superconformal symmetry of the
conformal mechanics model. In Section 4 we analyse the double
nonlinear superconformal symmetry of superconformal mechanics. In
Section 5 we discuss some problems to be interesting for further
investigation.

\section{Hidden superconformal symmetry of a free particle
on a line}

Let us start with a free unit mass particle on a line, which
is
given by the Hamiltonian
\begin{equation}\label{H0}
H=-\frac{1}{2}\,\frac{d^2}{dx^2}.
\end{equation}
A linear momentum $p=-i\frac{d}{dx}$ is  an integral of
motion
 satisfying the relation
\begin{equation}\label{p2H}
p^2=2H.
\end{equation}
As a consequence, every energy level with $E>0$ is doubly
degenerated: the two non-normalizable eigenstates
$\psi_{E,+}(x)=C_+\cos \sqrt{2E}\,x$ and
$\psi_{E,-}(x)=C_-\sin\sqrt{2E}\,x$ belong to it, while a
nondegenerate state $\psi_{0,+}=C_0=const$ corresponds to
$E=0$.
Due to the  energy levels structure, this elementary pure
bosonic
system possesses a hidden  $N=1$ supersymmetry in exact, not
spontaneously broken, phase. Indeed, take the reflection
operator
$R$,  $R\psi(x)=\psi(-x)$, satisfying the relations
$\{R,x\}=\{R,p\}=0$, $R^2=1$, where $\{.,.\}$ is an
anticommutator. Then the operators $Q_a$, $a=1,2$,
\begin{equation}\label{Q12}
Q_1=\frac{1}{\sqrt{2}}\,p,\qquad Q_2=iRQ_1,
\end{equation}
can be identified as Hermitian supercharges,
\begin{equation}\label{QQH}
\{Q_a,Q_b\}=2\delta_{ab}H,\qquad [Q_a,H]=0.
\end{equation} Their linear
combinations, $Q_\pm:=\frac{1}{\sqrt{2}}\,(Q_1\pm i
Q_2)=-i\frac{d}{dx}\cdot \frac{1}{2}(1\pm R)$, provide us
with
Hermitian conjugate nilpotent supercharges, $Q_+^2=Q_-^2=0$,
$\{Q_+,Q_-\}=2H$, which mutually transform the even,
$\psi_{E,+}$,
and the odd, $\psi_{E,-}$, eigenstates with $E>0$, and
annihilate
the ground state $\psi_{0,+}$. In this construction the
reflection
plays a role of a grading operator, which identifies the $H$
as an
even generator and the $Q$'s as odd generators of the $N=1$
superalgebra (\ref{QQH}).

The described  hidden (``bosonized" \cite{Bos,ParaP,Rau})
$N=1$
supersymmetry can be extended to the $osp(2|2)$
superconformal
symmetry by supplying the set of the integrals $H$ and $Q_a$
with
the \emph{dynamical} odd,
$$
S_1=\frac{1}{\sqrt{2}}\,X,\qquad S_2=iRS_1,
$$
and even, $D=\frac{1}{4}\{X,p\}$, $K=\frac{1}{2}X^2$, integrals,
where $X:=x-tp$. The $D$ and $K$ are presented equivalently as
\begin{equation}\label{DK}
D=\frac{1}{4}\{x,p\}-tH,\qquad K=\frac{1}{2}x^2-2tD-t^2H.
\end{equation}
The dynamical integrals of motion satisfy the equation of
the form
\begin{equation}\label{dynI}
\frac{d}{dt}I=\frac{\partial I}{\partial t}-i[I,H]=0.
\end{equation}
The $Q_a$, $S_a$, $H$, $D$, $K$  and the operator
\begin{equation}\label{Sig}
\Sigma=-\frac{1}{2}R
\end{equation}
satisfy the
$osp(2|2)$ superalgebra given by
 the nontrivial (anti)commutation relations
\begin{eqnarray}
&[H,K]=-2iD,\qquad [D,H]=iH,\qquad [D,K]=-iK,&\label{HDK}\\
&\{Q_a,Q_b\}=2\delta_{ab}H,\qquad \{S_a,S_b\}=2\delta_{ab}K, \qquad
\{S_a,Q_b\}=2\delta_{ab}D-\epsilon_{ab}\Sigma,&
\label{scon0}\\
&[H,S_a]=-iQ_a,\qquad [K,Q_a]=iS_a,\qquad
[D,Q_a]=\frac{i}{2}Q_a,\qquad
[D,S_a]=-\frac{i}{2}S_a,&\label{DQS}\\
&[\Sigma,Q_a]=i\epsilon_{ab}Q_b,\qquad
[\Sigma,S_a]=i\epsilon_{ab}S_b,&\label{SigQS}
\end{eqnarray}
in which the even generators form the bosonic
$so(1,2)\oplus u(1)$
subalgebra, while the odd generators form a pair of its
spin-$\frac{1}{2}$ representations.

\section{Hidden superconformal symmetry of conformal
mechanics}

Let us turn now to the conformal mechanics model \cite{AFF}
described by the Hamiltonian
\begin{equation}\label{Hconf}
H_\alpha=\frac{1}{2}\left(-\frac{d^2}{dx^2}+\frac{g(\alpha
)}{x^2}\right),\qquad
{\rm where}\qquad g(\alpha):=\alpha(\alpha+1),
\end{equation}
$0<x<\infty$, and it is assumed that the wave functions are
subjected to the boundary conditions $\psi(x)\rightarrow 0$,
$\psi'(x)\rightarrow 0$ for $x\rightarrow 0$, that we write
in the
form
\begin{equation}\label{BC}
\psi(0)=0,\qquad \psi'(0)=0.
\end{equation}
Formally, for  $g=0$ the model is reduced to a free particle
on a
half-line. However, for $-\frac{1}{4}\leq g < \frac{3}{4}$
the
operator (\ref{Hconf}) is not essentially self-adjoint (and
therefore cannot play the role of a Hamiltonian, for the
details
see refs. \cite{Me,Narn,Basu,Falom1,Govi}). On the other
hand,
Hamiltonian (\ref{Hconf}) with $g \geq \frac{3}{4}$ is
essentially
self-adjoint, and in what follows we shall assume that the
parameter $\alpha$ takes the values corresponding to the
latter
case.

 We shall  show that analogously to the model of the free
 particle
 on the line, for integer values of the parameter
 $\alpha=l$,
$l=1,2,\ldots$, (or, equivalently, for $-\alpha=2,3,\ldots$)
the
conformal mechanics model is described by the nonlinear
superconformal symmetry $osp(2|2)_{2l+1}$. It generalizes
the
hidden superconformal symmetry of the free particle, and is
produced by the even $so(1,2)\oplus u(1)$ generators $H_l$,
$D_l$,
$K_l$ and $\Sigma$, and by the set of odd operators
$S^+_{n,m}$,
$S^-_{n,m}$, $n=2l+1$, $m=0,1,\ldots, n-1$. The odd
generators
constitute the pair of spin-$(l+\frac{1}{2})$
representations of
the $so(1,2)\oplus u(1)$, and anticommute for order $2l+1$
polynomials of the even generators. As we shall see, it is
this
hidden supersymmetry of the purely bosonic conformal
mechanics
model (\ref{Hconf}), (\ref{BC}) that explains the origin and
nature of the double superconformal symmetry in the
superconformal
mechanics at certain values of the boson-fermion coupling
parameter \cite{LeiP}. On the other hand, it will be shown
that
unlike the case of the model (\ref{H0}), the hidden
superconformal
nonlinear symmetry of the conformal mechanics model is of
the
broken nature.

System (\ref{Hconf}) has two dynamical integrals of motion,
$D_\alpha$ and $K_\alpha$, given by Eq. (\ref{DK}) with
$H$ changed for (\ref{Hconf}). Together with
$H_\alpha$,
they form the $so(1,2)$ algebra (\ref{HDK}). Define the
operator
\begin{equation}\label{Da}
\nabla_\gamma=\frac{d}{dx}+\frac{\gamma}{x},\qquad
\nabla_\gamma^\dagger=-\nabla_{-\gamma},
\end{equation}
where $\gamma$ is a real parameter,
 and write down the relation
\begin{eqnarray}
\begin{array}{ccc}
  -2H_\alpha= &  &  \\
   & \nabla_{\alpha+1} \nabla_{-(\alpha+1)}=\nabla_{-\alpha}
   \nabla_{\alpha}&  \\
  &  & =-2H_{-(\alpha+1)}, \\
\end{array}
\label{factor}
\end{eqnarray}
which is valid for Hamiltonian (\ref{Hconf})  and is based
on the
elementary equality $g(\alpha)=g(-\alpha-1)$. In terms of
operator
(\ref{Da}), construct the order $n$ differential operator
\begin{equation}\label{Pan}
{\cal P}_{\gamma,n}:=(-i)^n \nabla_{\gamma
-n+1}\nabla_{\gamma-n+2}\ldots \nabla_{\gamma-1}\nabla_{
\gamma}.
\end{equation}
Due to definition (\ref{Da}), operator (\ref{Pan}) is
Hermitian in the case
\begin{equation}\label{an}
    \gamma=\frac{1}{2}(n-1),
\end{equation}
i.e. when parameter $\gamma$ takes integer or half-integer
values.
When $n$ takes an even value, $n=2l$, $l=1,2\ldots$, and in
accordance with (\ref{an})  $\gamma$ is half-integer,
(\ref{Pan})
takes the form
\begin{equation}
{\cal P}_{l-\frac{1}{2},2l}=(-1)^l \nabla_{-(l-\frac{1}{2})}
\ldots
\nabla_{-\frac{1}{2}}\nabla_{\frac{1}{2}}\ldots
\nabla_{l-\frac{1}{2}}. \label{Panev}
\end{equation}
Making use of relation (\ref{factor}) starting from the
center, we
obtain the chain of equalities
$$
...\nabla_{-\frac{3}{2}}\nabla_{-\frac{1}{2}}\nabla_{\frac{1
}{2}}\nabla_{\frac{3}{2}}...=
...\nabla_{-\frac{3}{2}}\nabla_{\frac{3}{2}}\nabla_{-\frac{3
}{2}}\nabla_{\frac{3}{2}}...=
...\nabla_{\frac{5}{2}}\nabla_{-\frac{5}{2}}\nabla_{\frac{5
}{2}}\nabla_{-\frac{5}{2}}...=\ldots,
$$
and find that operator (\ref{Panev}) is reduced to the $l$-th
order
of the operator $H_{l-\frac{1}{2}}$:
\begin{equation}\label{PHl}
{\cal P}_{l-\frac{1}{2},2l}=(2H_{l-\frac{1}{2}})^l.
\end{equation}
 Therefore, the
Hermitian operator (\ref{Panev}) is an integral of motion
for the
system (\ref{Hconf}) with $\alpha=l-\frac{1}{2}$, but it is
reduced to the $l$-th order of the Hamiltonian itself.

Consider now the case of the odd  $n=2l+1$  and integer
$\gamma=l$. Then, we have the  order $(2l+1)$ differential
operator (\ref{Pan}) of the form
\begin{equation}\label{Panodd}
{\cal P}_{l,2l+1}:=(-i)^{2l+1}
\nabla_{-l}\nabla_{-l+1}\ldots\nabla_0\ldots
\nabla_{l-1}\nabla_{l}.
\end{equation}
Let us show that it is an integral of motion for system
(\ref{Hconf}) with $\alpha=l$. First, this is so for $l=0$
when
Hamiltonian (\ref{Hconf}) is reduced to the formal free
particle
Hamiltonian (\ref{H0}) (see the comment on self-adjointness
above,
which, however, is not important at the moment), and first
order
operator (\ref{Panodd}) is reduced to the momentum operator.
In a
generic case, with taking into account Eq. (\ref{factor}) we
have
$$
[{\cal P}_{l,2l+1},H_{l}]=-\frac{1}{2}[{\cal
P}_{l,2l+1},\nabla_{-l}\nabla_{l}]=\frac{1}{2}\nabla_{-l}[{\cal
P}_{l-1,2l-1},\nabla_l\nabla_{-l}]\nabla_l=-\nabla_{-l}[{\cal
P}_{l-1,2l-1},H_{l-1}]\nabla_l.
$$
Since $[{\cal P}_{0,1},H_0]=0$, by induction we conclude
that
$[{\cal P}_{l,2l+1},H_{l}]=0$ for any $l=0,1,2,\ldots$. A
simple
algebraic calculation with repeated application of relation
(\ref{factor}) shows that the integral (\ref{Panodd})
satisfies
the relation
\begin{equation}\label{Panod2}
({\cal P}_{l,2l+1})^2=(2H_{l})^{2l+1},
\end{equation}
cf. the free particle relation (\ref{p2H}) corresponding to
$l=0$.

In what follows, remembering the problem of self-adjointness
for
operator (\ref{Hconf}), we shall assume that
\begin{equation}\label{al}
\alpha=l,\qquad l=1,2,\ldots.
\end{equation}

Let us denote $n=2l+1$ and $S_{n,0}:={\cal P}_{l,2l+1}$. The
commutator of any two dynamical integrals is also a
dynamical
integral. Then we find that the subsequent commutation of
the
integral $S_{n,0}$ with dynamical integral $K_{l}$ produces
a set
of the new $n-1$ dynamical integrals in accordance with
relation
\begin{equation}\label{KS}
[K_{l},S_{n,m}]=-(n-m)S_{n,m+1},\qquad m=0,\ldots,n-1.
\end{equation}
The $n$-times repeated commutator of $K_{l}$ with $S_{n,0}$
produces finally zero and we obtain the finite chain of
dynamical
integrals $S_{n,m}$, $m=1,\ldots n-1$ in addition to the
integral
$S_{n,0}$. These additional integrals, like $S_{n,0}$, are
also
the order $n=2l+1$ differential operators, and all together
they
explicitly can be presented in the form
\begin{equation}\label{Snm}
S_{n,m}=(-i)^{n} (x+it\nabla_{-l})(x+it \nabla_{-l+1})\ldots
(x+it\nabla_{-l+(m-1)})\nabla_{-l+m}\ldots \nabla_{l},\qquad
m=0,\ldots, n-1.
\end{equation}
To get this explicit form, we have used, in particular,  the
equality $ \nabla_{\gamma}x=x\nabla_{\gamma+1}. $ The
$S_{n,m}$
satisfies the relations $S_{n,m}^\dagger=(-1)^mS_{n,m}$ and
$$
\frac{\partial S_{n,m}}{\partial t}=imS_{n,m-1}.
$$
Therefore, being the dynamical integral of motion, the
$S_{n,m}$
satisfies the commutation relation
\begin{equation}\label{HS}
[H_{l}, S_{n,m}]=-mS_{n,m-1}.
\end{equation}
We have also the relation
\begin{equation}\label{DS}
[D_{l},S_{n,m}]=i\left(\frac{n}{2}-m\right)S_{n,m}.
\end{equation}
According to  (\ref{KS}), (\ref{HS}) and (\ref{DS}), the set of the
operators $S_{n,m}$ forms the $so(1,2)$ spin-$\frac{n}{2}$
representation. All the $S_{n,m}$ are the order $n=2l+1$
differential operators in $x$, and so, anticommute with the
reflection operator $R$, while the $so(1,2)$ generators are even
operators commuting with $R$. Then, as in the free particle case,
one can extend the set of odd operators $S_{n,m}$ with the set of
odd operators $iRS_{n,m}$, which satisfy the commutation relations
with the $so(1,2)$ generators exactly of the same form as $S_{n,m}$.
To distinguish these two sets of odd operators, we, again, add  the
$u(1)$ generator (\ref{Sig}) to the set of even operators. All this
set of integrals forms the nonlinear superconformal algebra
$osp(2|2)_{2l+1}$ given by nontrivial relations of the form
(\ref{HDK}) and
\begin{equation}
[\Sigma,S{}^\pm_{n,m}]=\pm  S{}^\pm_{n,m}, \quad [D_l,
S{}^\pm_{n,m}]=i\left(\frac{n}{2}-m\right)  S{}^\pm_{n,m},
\label{sds}
\end{equation}
\begin{equation}
[H_l, S{}^\pm_{n,m}]=\mp m S{}^\pm_{n,m-1},\quad [ K_l,
S{}^\pm_{n,m}]=\mp (n-m)  S{}^\pm_{n,m+1}, \label{hks}
\end{equation}
\begin{eqnarray}
\{ S{}^+_{n,m}, S{}^-_{n,m'}\}=
P_{2l+1}^{m,m'}(H_l,K_l,D_l,\Sigma) \label{ssquant}
\end{eqnarray}
where $S^+_{n,m}=\frac{1}{2}(1- R)S_{n,m}$,
$S^-_{n,m}=(-1)^m\frac{1}{2}(1 + R)S_{n,m}=(S^+_{n,m})^\dagger$, and
$P_{2l+1}^{m,m'}$ is an order $n=2l+1$ polynomial of its arguments,
whose explicit form is not important for us here (see ref.
\cite{AnaP}).

We have found the odd integrals of motion not taking into
account
the boundary condition (\ref{BC}). To clarify this aspect,
we turn
to the spectral problem for Hamiltonian (\ref{Hconf}).
Having in
mind that for $g\geq  \frac{3}{4}$ the system has no states
with
$E\leq 0$,  we introduce the notations
\begin{equation}\label{nota}
k=\sqrt{2E},\qquad z=kx,\qquad \psi(x)=\sqrt{z}\, u(z),
\qquad
\nu=\alpha+\frac{1}{2},
\end{equation}
implying that $E>0$. Then the spectral equation $(H_\alpha
-E)\psi(x)=0$ is reduced to the Bessel equation
\begin{equation}\label{BE}
z^2\frac{d^2 u}{dz^2}+z\frac{du}{dz}+(z^2-\nu^2)u=0.
\end{equation}
For non-integer values of $\nu$ (in this case $\alpha$ is
not
half-integer, that includes (\ref{al})), the general
solution of
Eq. (\ref{BE}) is $u(z)=AJ_\nu(z)+BJ_{-\nu}(z)$, where
$J_\nu(z)$
is Bessel function, and $A$, $B$ are some constants. Bessel
functions satisfy the recursive differential relations
\begin{equation}\label{JJ}
\nabla_{-\nu}\,J_\nu(z)=-J_{\nu+1}(z),\qquad
\nabla_{\nu}\,J_\nu(z)=J_{\nu-1}(z),
\end{equation}
where $\nabla_\nu$ is the first order differential operator
given
by Eq. (\ref{Da}) with $x$ changed for $z$. Using repeatedly
the
second relation, we get
\begin{equation}\label{nablan}
\tilde{\cal P}_{\nu,n+1}\,J_\nu(z)=J_{\nu-(n+1)}(z),\qquad
\tilde{\cal
P}_{\nu,n+1}:=\nabla_{\nu-n}\nabla_{\nu-n+1}\ldots\nabla_\nu
.
\end{equation}
When $\nu=l+\frac{1}{2}$ and $n=2l$, the operator $\tilde{\cal
P}_{l+\frac{1}{2},2l+1}$ transforms the solution
$J_{l+\frac{1}{2}}(z)$ into independent solution
$J_{-(l+\frac{1}{2})}(z)$ of the same equation (\ref{BE}). Since
$J_\nu(z)\sim z^\nu$ for $z\sim 0$, from the two solutions
$J_{l+\frac{1}{2}}(z)$ and $J_{-(l+\frac{1}{2})}(z)$ of the same
equation (\ref{BE}) with $\nu^2=(l+\frac{1}{2})^2$ only the first
one satisfies the boundary condition (\ref{BC}) corresponding to the
conformal mechanics model. Remembering relations (\ref{nota}), we
find that the action of the operator $\tilde{\cal
P}_{l+\frac{1}{2},2l+1}$ on the function $u(z)$ up to a numerical
coefficient  is reduced to the action of the supercharge $S_{n,0}$
on the corresponding function $\psi(x)$. This means that when the
integral $S_{n,0}$ acts on a physical state, which is an eigenstate
of the Hamiltonian $H_l$ satisfying boundary condition (\ref{BC}),
it produces a state which formally still is  an eigenstate of the
same differential operator $H_l$ but does not satisfy the boundary
condition at the origin. This picture is somewhat reminiscent to the
the spontaneously broken supersymmetry, in which a zero energy state
being non-normalisable does not satisfy the boundary condition at
infinity. On the other hand, here, unlike the usual spontaneously
broken supersymmetry, we have no pairing of the states even in a
part of the spectrum. Because of this reason the hidden symmetry of
the conformal mechanics can be called a virtual superconformal
symmetry.

 Note also that taking in (\ref{nablan})
$n=2l+1$, $\nu=l+1$, $l=0,1,\ldots$, and using the earlier
observed relation (\ref{PHl}), we reproduce the well known
identity
\begin{equation}\label{Jint}
J_{-l}(z)=(-1)^lJ_l(z).
\end{equation}

\section{Double superconformal symmetry of superconformal
mechanics}

The Hamiltonian of the system possessing superconformal
symmetry
of order $n$ \cite{LeiP,AnaP} can be presented in the form
\begin{equation}\label{Hna}
H_{n,\alpha}=\left(%
\begin{array}{cc}
  H_{\alpha -n} & 0 \\
  0 & H_\alpha \\
\end{array}%
\right),
\end{equation}
where the upper and lower Hamiltonian operators are of the
form
(\ref{Hconf}), and the upper, $\psi^+$, and lower, $\psi^-$,
components of the state $\Psi^T=(\psi^+,\psi^-)$, are
subjected to
the boundary condition (\ref{BC}). For the sake of
definiteness we
shall assume that the parameter $\alpha$ is non-negative.
The set
of odd generators of nonlinear superconformal symmetry
associated
with Hamiltonian (\ref{Hna}) has the structure similar to
(\ref{Snm}),
\begin{equation}\label{San}
S^+_{n,m;\alpha}:=(x+it\nabla_{\alpha
-n+1})(x+it\nabla_{\alpha-n+2})\ldots (x+it\nabla_{\alpha-n+l}){\cal
P}_{\alpha,n-l}\sigma_+,
\end{equation}
where $\sigma_+=\frac{1}{2}(\sigma_1+i\sigma_2)$, and operator
${\cal P}_{\alpha,n-l}$ is defined by Eq. (\ref{Pan}). These odd
supercharges together with conjugate operators anticommute for order
$n$ polynomials in even generators $H_{n,\alpha}$, $K_{n,\alpha}$,
$D_{n,\alpha}$ and $\Sigma=\frac{1}{2}\sigma_3$ (for the details see
refs. \cite{LeiP,AnaP}). In the case $n=1$ the system (\ref{Hna})
corresponds to the superconformal mechanics model \cite{AP,FR}
possessing the $osp(2|2)$ superconformal symmetry of the form
(\ref{HDK})--(\ref{SigQS}).

Before we pass over to the discussion of the double superconformal
symmetry, let us note that applying the results of the previous
section, we find that when the parameters $\alpha$ and $n$ satisfy
the relation $\alpha>n$, we have the order $n$ unbroken
superconformal symmetry. In this case, in particular, when the
supercharges $S^\pm_{n,0;\alpha}$ commuting with the Hamiltonian act
on the eigenstates of (\ref{Hna}) satisfying boundary condition
(\ref{BC}), they mutually transform these eigenstates. For
$\alpha=n$, there appears the problem with self-adjointness of the
supersymmetric Hamiltonian (\ref{Hna}). Finally, when the parameters
satisfy the relation $0<\alpha<n$ and $\alpha\neq l+\frac{1}{2}$,
$l=0,1,\ldots,n-1$, the order $n$ superconformal symmetry is broken
in the same way as the hidden symmetry discussed in the previous
section. Acting on the Hamiltonian eigenstates being proportional to
$J_{\alpha+\frac{1}{2}}$ or $J_{n-\alpha+\frac{1}{2}}$, the
$S^\pm_{n,0;\alpha}$ will produce the states violating boundary
condition (\ref{BC}) except for the case when $\alpha=
l+\frac{1}{2}$, $l=0,1,\ldots,n-1$. In the last case due to relation
of the form (\ref{Jint}) we have the unbroken nonlinear
superconformal symmetry $osp(2|2)_n$.

Now we turn to the double superconformal symmetry. In ref.
\cite{LeiP} it was observed that when the parameter $\alpha$
takes
an integer value, the system (\ref{Hna}) can simultaneously
be
characterized by the two nonlinear superconformal symmetries
of
the  orders shifted in the odd number. More specifically,
when
$\alpha=n+p$, $p=1,2,\ldots$, the system (\ref{Hna}) is
characterized also, in addition to the nonlinear
superconformal
symmetry of the order $n$, by the superconformal symmetry of
the
order $n'=n+2p+1$. Let us show that this second
supersymmetry
really is of the broken, virtual  nature. Indeed, the
supercharge
$S^+_{n+2p+1,0;n+p}$ constructed in accordance with Eq.
(\ref{San}), commutes with Hamiltonian (\ref{Hna}) and
anticommutes with its conjugate operator for the operator
$(H_{n,n+p})^{n+2p+1}$. It can be presented in the form
\begin{equation}\label{Svirt}
S^+_{n+2p+1,0;n+p}=(\nabla_{-p}\ldots
\nabla_p)S^+_{n,0;n+p}=i^{2p+1}{\cal
P}_{p,2p+1}S^+_{n,0;n+p}\, .
\end{equation}
The operator ${\cal P}_{p,2p+1}$, as we have seen, commutes
with
the conformal mechanics model Hamiltonian $H_{p}$, but
acting on
an eigenstate of the latter which satisfies boundary
condition
(\ref{BC}), it produces a state violating (\ref{BC}). So, we
conclude that the higher order symmetry of the system with
double
superconformal symmetry (like in the case $0<\alpha<n$,
$\alpha\neq l+\frac{1}{2}$,
mentioned
above) has the same  broken, virtual nature as a hidden
nonlinear
superconformal symmetry of the conformal mechanics model
with
integer parameter (\ref{al}).

For the sake of completeness, we note that the double
nonlinear
superconformal symmetry with supersymmetry orders shifted in
an
even number is trivial  \cite{LeiP}: in this case
the parameter $\alpha$ takes a half-integer value and
the structure of the
higher
order odd generators is reduced to the corresponding lower
order
odd generators multiplied by monomials in bosonic generators
due
to relations of the form  (\ref{PHl}).

\section{Discussion and outlook}
We have seen that the breaking of hidden superconformal
symmetry
in conformal mechanics model  is complete in the sense that
no
energy eigenstates pairing remains in the system.
 It would be interesting to clarify whether there exist
 some systems in which the supersymmetry would be broken via
 violation of a boundary condition at the finite extremum
 only for the part of the
Hamiltonian  eigenstates. If so, it could provide us with
the
supersymmetry breaking mechanism different from the usual
mechanism of spontaneous supersymmetry breaking associated
with
the lost of normalisability of the ground state via
violation of
boundary condition at infinity.

We have shown that the higher order symmetry of the double
nonlinear superconformal symmetry of superconformal
mechanics has
the same, broken nature as a hidden superconformal symmetry
of the
conformal mechanics model. This is because  the
corresponding
higher order supercharges have the structure of the lower
order
superconformal symmetry generators multiplied by  the
differential
operators generating the described hidden superconformal
symmetry
of the corresponding bosonic subsystem. As a result, the
higher
order superconformal symmetry odd generators acting on the
Hamiltonian eigenstates produce the states which violate the
boundary condition at the origin.

Other systems with double nonlinear supersymmetry were found outside
the context of conformal symmetry \cite{And,AS}. However, potentials
of such systems reveal a structure similar to the potential of the
conformal mechanics: they are singular at the origin. Moreover under
appropriately taken parameter limit, they are reduced to the $1/x^2$
potential. Therefore, other known systems with double supersymmetry
belong to the class of the systems to which the conformal and
superconformal mechanics model do belong: they are the systems
formulated on a half-line, and so, include the boundary condition at
the origin as a defining ingredient. By analogy, one could expect
that the higher order supersymmetry in such systems should also be
of the broken, virtual nature. Note here that independently from the
nonlinear supersymmetry  context \cite{AIS,ParaP,KlP,ASaT}, the
supersymmetry breaking in the systems with singular potentials was
discussed recently in \cite{Falom2}. We are going to present the
results of investigation of such systems elsewhere.


\vskip 0.4cm\noindent {\bf Acknowledgements}. MP is indebted to the
University of Valladolid for hospitality extended to him, and thanks
H. Falomir and E. Ivanov for discussions. This work has been
partially supported by the FONDECYT-Chile (projects 1050001 and
7050046), by the DGI of the Ministerio de Educaci\'on y Ciencia of
Spain (project BMF2002-02000), by the FEDER Programme of the
European Community, and by the Junta de Castilla y
Le\'on (project
VA013C05).



\end{document}